\documentclass[aps,eqsecnum,nofootinbib,showpacs]{revtex4}
\usepackage[dvips]{graphicx}  
\usepackage{amsmath,amsfonts}
\usepackage{bm}
\usepackage{color}  
\begin{document}
\title{ A Solvable Model for Fermion Masses \\
on a Warped 6D World with the Extra 2D Sphere
}
\author{Akira Kokado}
\email{kokado@kobe-kiu.ac.jp}
\affiliation{Kobe International University, Kobe 658-0032, Japan}
\author{Takesi Saito}
\email{tsaito@k7.dion.ne.jp}
\affiliation{Department of Physics, Kwansei Gakuin University,
Sanda 669-1337, Japan}
\date{\today}
\begin{abstract}
In a warped 6D world with an extra 2-dimensional sphere, we propose an exactly solvable model for fermion masses with zero mode. The warp factor is given by  $\phi (\theta ,\varphi )=\sin{\theta }\cos{\varphi }$, which is a solution to the 6D Einstein equation with the bulk cosmological constant $\Lambda $ and the energy-momentum tensor of the bulk matter fields. Our model provides another possibility of obtaining fermion zero mode, rather than traditional model based on Dirac's monopole.
\end{abstract}

\maketitle
\section{Introduction}\label{sec:intro}
%
%
The 6-dimensional space is particularly interested in unified theories in higher-dimensions. The extra 2-dimensional compact space generates some useful gauge symmetries for various fields \cite{ref:Camporesi, ref:Abrikosov}.\\
\indent In this article we specially confine ourselves to the problem of fermionic masses when the extra 2-dimensional surface is a sphere. In the case where the 4-dimensional space is Minkowskian, we encounter a serious theorem that there is no zero mode in a 4-dimensional fermionic field \cite{ref:Camporesi, ref:Abrikosov}. This is well known in the spectrum of the Dirac operator on the 2D sphere where we have the SU(2) symmetry. This theorem can be generalized to any internal space with a positive curvature. Since it is desirable that we have a zero mode in unified theories at least at the first symmetric stage, this theorem is unwelcome. To overcome this difficulty it has been considered to introduce a gauge field. When the gauge field has Dirac's monopole, we get the zero mode of fermionic fields  \cite{ref:Fujii, ref:Cheng}.\\
\indent As another possibility of obtaining fermion zero mode, we propose an exactly solvable model in the warped 6D world with the extra 2D sphere. The line element of this model is
\begin{align}
 &  ds^2 = g_{AB} dx^{A} dx^{B} =\phi ^{2}(\theta , \varphi ) \eta _{\mu \nu }dx^{\mu }dx^{\nu } - a^2  \big(d\theta ^2 + \sin^{2}{\theta } d\varphi ^{2} \big) ~,
 \label{eq:metric}
\end{align}
where the 4-dimensional metric,  $\eta _{\mu \nu }$, has the signature (+,-,-,-), and the extra 2-dimensional surface is a sphere with a constant radius $a$ and the two spherical angles $x^{5}=\theta $ and $x^{6}=\varphi$.  In the following we assume $a<<1$, in order to make KK modes negligible in the present-time energy scale. The warp factor is given by 
\begin{align}
 &  \phi (\theta , \varphi ) = \sin{\theta } \cos{\varphi },
  \label{eq:warp_facter}
\end{align}
which is a solution to Einstein's equation with the bulk cosmological constant $\Lambda $ and the energy-momentum tensor of the bulk matter fields. We solve the 6D Dirac equation with a 6D zero-mass in this 6D warped background (\ref{eq:metric}). We then find a 4D fermion mass formula and also a zero mode solution.
\\
\indent There are many other attempts to obtain fermion masses in warped extra dimensions. In Refs~\cite{ref:Gogberashvili1, ref:Aguilar, ref:Gogberashvili2, ref:Parameswaran, ref:Williams, ref:SenGupta, ref:Das, ref:Koley, ref:Choudhury, ref:Lawrance, ref:Guo, ref:Biggio} there are no symmetries like SU(2) in extra dimensions, so that the fermion zero mode is included. In Ref.~\cite{ref:Steinacker} the 4D fermion coupled with a SU(N) gauge field has been considered. The gauge field spontaneously generates fuzzy extra dimensions, which yield a theory on $M^4\times S^2$, then resulting a fuzzy zero-mode. \\
\indent In our case, at first sight the extra 2D sphere seems to be SU(2)-symmetric. However, the warp factor (\ref{eq:warp_facter}) has zeros at $\varphi =\pm\pi /2$ and $\theta =0,\pi $, so that internal wave functions violate the SU(2) symmetry there. This is a reason why our fermion has a zero mode.\\ 
\indent In Sec.\ref{sec:2} the warp factor (\ref{eq:warp_facter}) is derived. In Sec.\ref{sec:3} the Dirac equation in 6D warped background is introduced.  
In Sec.\ref{sec:4} the Dirac equation is solved in the method of separable variables to obtain the 4D fermion mass formula. The final section is devoted to concluding remarks. The Appendix is prepared for constraints coming from the back reaction of the fermion.
  
\section{The warp factor}\label{sec:2}
The action of the gravitational system in six dimensions can be written as 
\begin{align}
 &  I = -\frac{1}{2\kappa_{6} ^2} \int d^6x \sqrt{-g}\big( R + 2\Lambda \big) ~,
  \label{eq:Action}
\end{align}
where $\kappa_{6} ^2=8\pi G_{N}$ is the six dimensional Newton constant and $\Lambda $ is the bulk cosmological constant.  When we have a stress-energy tensor $T_{AB}$ in the bulk, Einstein equations become
\begin{align}
 &  R_{AB} - \frac{1}{2}g_{AB} R = \kappa_{6} ^2 \big( \Lambda g_{AB} + T_{AB}\big)~.
  \label{eq:Einstein_Eq}
\end{align}
Capital Latin indices run over $A, B, \cdots=0,1,2,3,5,6$. \\
\indent We look for solutions of Eq.(\ref{eq:Einstein_Eq}) with the ansatz for the stress-energy tensor of the bulk matter fields \cite{ref:Gogberashvili1, ref:Kanti, ref:Oda}: 
\begin{align}
 & T_{\mu \nu }=-g_{\mu \nu }E(\theta ,\varphi ), \quad T_{ij}=-g_{ij}P(\theta ,\varphi ), 
 \quad T_{i\mu }=0~,
  \label{eq:ansatz} 
\end{align}
The results are obtained in the following: From Eq.(\ref{eq:Einstein_Eq}) we have four equations as
\begin{align}
 & \frac{3}{\phi ^2}\Big(\frac{\partial \phi }{\partial \theta }\Big)^2 + \frac{3}{\phi}\frac{\partial^2 \phi }{\partial \theta ^2} + \frac{3 \cos{\theta}}{\phi \sin{\theta}}\frac{\partial \phi }{\partial \theta } 
 + \frac{3}{\phi ^2 \sin^2{\theta}}\Big(\frac{\partial \phi }{\partial \varphi }\Big)^2 + \frac{3}{\phi \sin^2{\theta}}\frac{\partial ^2 \phi }{\partial \varphi ^2} - 1  
 =  \kappa_{6} ^2 a^2 \big(E - \Lambda \big)~,
 \label{eq:eq_mu_nu} \\
&  \frac{6}{\phi ^2 \sin^2{\theta}}\Big(\frac{\partial \phi }{\partial \varphi }\Big)^2 + \frac{4}{\phi \sin^2{\theta}}\frac{\partial ^2 \phi }{\partial \varphi ^2} 
 + \frac{4 \cos{\theta}}{\phi \sin{\theta}}\frac{\partial \phi }{\partial \theta } + \frac{6}{\phi ^2}\Big(\frac{\partial \phi }{\partial \theta }\Big)^2  
 = \kappa_{6} ^2 a^2 \big( P - \Lambda \big)~,
 \label{eq:eq_55} \\
& \frac{6}{\phi ^2 }\Big(\frac{\partial \phi }{\partial \varphi }\Big)^2  + \frac{6\sin^2{\theta}}{\phi ^2}\Big(\frac{\partial \phi }{\partial \theta }\Big)^2 + \frac{4\sin^2{\theta}}{\phi }\frac{\partial^2 \phi }{\partial \theta ^2}
 =  \kappa_{6} ^2 a^2 \sin^2{\theta}\big(P - \Lambda \big)~,
 \label{eq:eq_66} \\
 & \frac{\cos{\theta}}{\phi \sin{\theta}}\frac{\partial \phi }{\partial \varphi }  - \frac{1}{\phi}\frac{\partial^2 \phi }{\partial \theta \partial \varphi } = 0~.
 \label{eq:eq_56}
\end{align}
For such a solution that $\phi (\theta ,\varphi )=\Theta (\theta )\Phi (\varphi )$ takes a maximum value 1 at  $\theta =\pi /2$ and $\varphi =0$, the last equation is immediately solved as
\begin{align}
 & \Theta (\theta )= C \sin{\theta }~.
 \label{eq:solution_Theta} 
\end{align}
Substituting the result into the other three equations above, we get
\begin{align}
 & \kappa_{6} ^2 a^2 = \frac{10}{\Lambda }~,
\quad  \Phi (\varphi ) = \cos{\varphi }~, \quad C=1~,
 \label{eq:solution_Fai} \\
&  E(\theta ,\varphi )= \frac{3}{\kappa_{6} ^2 a^2 \phi ^2(\theta ,\varphi )}~,
 \quad
  P(\theta ,\varphi )= \frac{6}{\kappa_{6} ^2 a^2 \phi ^2(\theta ,\varphi )}~.
 \nonumber  
\end{align}
\indent To sum up we have

\begin{align}
 &\phi (\theta , \varphi ) = \sin{\theta } \cos{\varphi }~,
 \label{eq:fai2} \\
 & T_{AB} = \frac{3\Lambda }{10} \mbox{dial}\Big( -1, 1, 1, 1, \frac{2a^2}{\phi ^2}, \frac{2a^2\sin^2{\theta }}{\phi ^2}\Big)~.
 \label{eq:T_AB2} 
\end{align}
%
\section{The 6D Dirac equation} \label{sec:3}
We now consider the 6-dimensional massless Dirac equation with the metric (\ref{eq:metric}):
\begin{align}
 & ib_{\bar{A}}^{A} \Gamma ^{\bar{A}}D_{A}\Psi (x^{A}) = 0~,
 \label{eq3:Fermion_Eq}
\end{align}
where $D_{A}$ denote covariant derivatives, $\Gamma ^{\bar{A}}$ the 6-dimensional flat gamma matrices and $b_{A}^{\bar{A}}$ the sechsbein through the definition
\begin{align}
 & g_{AB} = \eta _{\bar{A}\bar{B}}b_{A}^{\bar{A}}b_{B}^{\bar{B}}~,
 \label{eq:def_g_AB}
\end{align}
where $\bar{A}, \bar{B}, \cdots$ are local Lorentz indices.\\
\indent In six dimensions a spinor
\begin{align}
 & \Psi (x^{A}) = \left(
   \begin{array}{c}
      \psi  \\
      \xi \\
   \end{array}
  \right)~,
 \label{eq:comp_Psi}
\end{align}
has eight components and is equivalent to a pair of 4-dimensional Dirac spinors, $\psi $ and $\xi $. We use the following representation of the flat ($8\times 8$) gamma-matrices
\begin{align}
 & \Gamma ^{\bar{\mu }} = \gamma ^{\bar{\mu }}\otimes \bm{1}~, \quad 
  \Gamma ^{\bar{\theta }} = i\gamma _{5}\otimes \tau _{1}~, \quad
 \Gamma ^{\bar{\theta }} = i\gamma _{5}\otimes \tau _{2}~, 
  \label{eq3:def_Gamma}
\end{align}
where $\tau _{i}$'s are Pauli matrices. In the following for simplicity we drop the bars of $\gamma $ matrices when no confusion will occur, i.e., $\gamma ^{\bar{\mu }}=\gamma ^{\mu }$, and $\gamma ^{5}=i\gamma ^{0}\gamma ^{1}\gamma ^{2}\gamma ^{3}$. They satisfy
\begin{align}
 & \big\{ \Gamma ^{\bar{A}}~, \Gamma ^{\bar{B}}~ \big\} = 2\eta ^{\bar{A}\bar{B}}~. 
 \label{eq3:commutation_relation}
\end{align}
with $\eta ^{\bar {A}\bar{B}}=(+1,-1,-1,-1,-1,-1)$.\\
\indent The sechsbein for our background metric (\ref{eq:metric}) is given by
\begin{align}
 & b_{\bar{A}}^{A}= \Big( \frac{1}{\phi }\delta _{\bar{\mu }}^{A}, \frac{1}{a}\delta _{\bar{\theta }}^{A}, \frac{1}{a\sin{\theta }}\delta _{\bar{\varphi }}^{A}\Big)~. 
 \label{eq:sechesbein} 
\end{align}
From the definition of standard spin-connections the non-vanishing components for them can be found
\begin{align}
 & \omega _{\nu }^{\bar{\theta }\bar{\mu }}= \frac{1}{a}\cos{\theta }\cos{\varphi }\delta ^{\bar{\mu }}_{\nu } ~,\quad \omega _{\nu }^{\bar{\varphi }\bar{\mu }}= -\frac{1}{a}\sin{\varphi } \delta _{\nu }^{\bar{\mu }}~, \quad \omega _{\varphi }^{\bar{\theta }\bar{\varphi }}=\cos{\theta }~.
 \label{eq:spinconnection}
\end{align}
The Dirac equation (\ref{eq3:Fermion_Eq}) then reduces to
\begin{align}
 & \Big[\frac{1}{\phi}\Gamma ^{\bar{\mu }}\partial _{\mu } + \frac{1}{a}\Gamma ^{\bar{\theta }}\big(\partial _{\theta } + \frac{1}{2}\cot{\theta } + 2\cot{\theta }\big) 
 + \frac{1}{a\sin{\theta }}\Gamma ^{\bar{\varphi }}\big(\partial _{\varphi } -2\tan{\varphi }\big) \Big]\Psi (x^A)=0~.
 \label{eq:Dirac_eq2}
\end{align}
If we put
\begin{align}
 & \Psi (x^{A}) \equiv  \frac{1}{\phi ^2\sqrt{\sin{\theta }}}\tilde {\Psi }(x^{\mu }, \theta , \varphi )~,
 \label{eq:Psi2}
\end{align} 
it follows that 
\begin{align}
 & \Big[\frac{1}{\phi}\Gamma ^{\bar{\mu }}\partial _{\mu } + \frac{1}{a}\Gamma ^{\bar{\theta }}\partial _{\theta } + \frac{1}{a\sin{\theta }}\Gamma ^{\bar{\varphi }}\partial _{\varphi } \Big]\tilde{\Psi }(x^A)=0~.
 \label{eq:Dirac_eq2a}
\end{align}
Define the operator $\hat {\nabla }$ by
\begin{align}
 & \hat {\nabla }\equiv  \big( \tau _{1} \partial _{\theta } + \tau _{2}\frac{1}{\sin{\theta }}\partial _{\varphi }\big)~.
 \label{eq:def_nabla} 
\end{align}
then Eq.(\ref{eq:Dirac_eq2a}) can be rewritten as
\begin{align}
 & \Big( ia \phi ^{-1} \gamma ^{\mu }\otimes \bm{1} \partial _{\mu } - \gamma _{5}\otimes \hat {\nabla }\Big) \tilde {\Psi }(x^{\mu }, \theta , \varphi )=0~.
 \label{eq:Dirac_eq3} 
\end{align}
\indent Let us expand $\tilde {\Psi }(x^{\mu }, \theta , \varphi )$ into eigenfunctions of the chiral operator $\gamma _{5}$  as follows:
\begin{align}
 & \tilde {\Psi }(x^{\mu }, \theta , \varphi ) = \psi _{R}(x^{\mu })f_{+}(\theta , \varphi ) + \psi _{L}(x^{\mu })f_{-}(\theta , \varphi )~,
 \label{eq:def_f}
\end{align}
where $\gamma _5 \psi _{R}(x^{\mu })=\psi _{R}(x^{\mu })$ and $\gamma _5 \psi _{L}(x^{\mu })=-\psi _{L}(x^{\mu })$. If we use the Dirac equation with mass $m$ in 4-dimensions,
\begin{align}
 & i \gamma ^{\mu }\partial _{\mu }\psi = m\psi ~,
 \label{eq:mass_eq}
\end{align}
or equivalently
\begin{align}
 & i \gamma ^{\mu }\partial _{\mu }\psi_{L} = m\psi_{R} ~, \quad i \gamma ^{\mu }\partial _{\mu }\psi_{R} = m\psi_{L}~,
 \label{eq:mass_eq2}
\end{align}
it follows that
\begin{align}
 & \hat{\nabla } f_{\pm}(\theta , \varphi ) = \pm \frac{ma}{\phi } f_{\mp} (\theta , \varphi )~.
 \label{eq:f_pm_Eq} 
\end{align}
These equations reduce to
\begin{align}
 & -i \hat{\nabla } g_{\pm}(\theta , \varphi ) = \pm \frac{ma}{\phi } g_{\pm} (\theta , \varphi )~,
 \label{eq:f_pm_eq2} \\ 
 & g_{\pm} \equiv  \frac{f_{+} \mp i f_{-}}{2}~.
 \nonumber
\end{align}
\indent Here it should be noted that there is no zero mode ($m=0$) for the 4D Dirac field $\psi (x^{\mu })$, when $\phi \equiv 1$. This is because the Dirac operator $\hat {\nabla }$ is just the Dirac operator on the sphere, its eigenvalues are known to be non-zero
\begin{align}
 & -i \hat{\nabla } g_{\lambda }(\theta , \varphi ) = \lambda  g_{\lambda } (\theta , \varphi )~,
 \label{eq:f_lambda_eq} 
\end{align}
with $\lambda =\pm 1, \pm 2, \cdots$ \cite{ref:Camporesi, ref:Abrikosov}. Here $g_{\lambda }(\theta , \varphi )$ is the spinor on the sphere. However, this is not always clear in the case of the 6D warped space. In the next section we would like to see how to fix the fermion mass.
%
\section{Fermion on the extra 2D sphere} \label{sec:4}
\subsection{A finite mass formula in 4-dimensions}
First we consider a case of $m \neq 0$, where $m$ is the fermion mass in 4D. The zero mode solution   will be considered in the separate subsection B.\\
\indent We would like to solve Eqs.(\ref{eq:f_pm_eq2}), i.e.,
\begin{align}
 & -i \Big(\tau _{1} \partial _{\theta } + \tau _{2} \frac{1}{\sin{\theta }}\partial _{\varphi }\Big) g_{\pm}(\theta , \varphi ) = \pm \frac{ma}{\sin{\theta }\cos{\varphi } } g_{\pm} (\theta , \varphi )~. 
 \label{eq:f_pm_eq}
\end{align}
This is a separable type of variables, so that we put
\begin{align}
 &  g_{\pm} (\theta , \varphi ) 
 \equiv  \alpha_{\pm} (\theta ) \left(
   \begin{array}{c}
      u_{\pm}(\varphi )  \\
      v_{\pm}(\varphi ) \\
   \end{array}
  \right)~,
 \label{eq:g_pm} 
\end{align}
Then we have 
\begin{align}
 & \partial _{\theta } \alpha (\theta ) =  \frac{C_{\pm}}{\sin{\theta } } \alpha (\theta ) ~,
 \label{eq:alpha_Eq}
\end{align}
and
\begin{align}
 & \big(\partial _{\varphi } -iC_{\pm}\big) u_{\pm}(\varphi ) = \pm \frac{ma}{\cos{\varphi } } v_{\pm}(\varphi ) ~,
 \label{eq:u_Eq} \\
& \big(\partial _{\varphi } + iC_{\pm}\big) v_{\pm}(\varphi ) = \mp \frac{ma}{\cos{\varphi } } u_{\pm}(\varphi ) ~,
 \label{eq:v_Eq}
\end{align}
where $C_{\pm}$ is an arbitrary constant. \\
\indent The first equation (\ref{eq:alpha_Eq}) can be solved generally as
\begin{align}
 & \alpha (\theta ) = \tan^{C_{\pm}}(\frac{\theta }{2}) ~.
 \label{eq:solution_alpha}
\end{align}
Here a normalization factor is absorbed into $u_{\pm}$ and $v_{\pm}$. \\ 
\indent From coupled equations (\ref{eq:u_Eq}) and (\ref{eq:v_Eq}) we have equations of the Schrodinger type, 
\begin{align}
 & \Big[\partial^2 _{\varphi } - \tan{\varphi } \partial _{\varphi } + iC_{\pm} \tan{\varphi } + C_{\pm}^2  + \frac{m^2a^2}{\cos^2{\varphi }}\Big] u_{\pm}(\varphi )  = 0~. 
 \label{eq:u_eq2}
\end{align}
$v_{\pm}(\varphi )$ are obtained by substituting the solutions $u_{\pm}(\varphi )$ into Eq. (\ref{eq:u_Eq}). \\
If we put
\begin{align}
 & u(\varphi )\equiv  e^{iC_{\pm}\varphi } h_{\pm}(z )~,
 \label{eq:def_u}
\end{align}
The equations (\ref{eq:u_eq2}) reduce to 
\begin{align}
 & \Big[\partial^2 _{\varphi } + (2iC_{\pm}- \tan{\varphi }) \partial _{\varphi } + \frac{m^2a^2}{\cos^2{\varphi }}\Big] h_{\pm}(\varphi )  = 0~. 
 \label{eq:u_eq3}
\end{align}
In order that the wave functions $u_{\pm}(\varphi )$ are periodic functions of $\varphi $ on the sphere, $C_{\pm}$  should take integer or half-integer values. If we introduce a variable $z$  defined by
\begin{align}
 & z  \equiv \frac{1+i\tan{\varphi }}{2}=\frac{e^{i\varphi }}{2\cos{\varphi }}~.
 \label{eq:def_zeta}
\end{align}
The equations (\ref{eq:u_eq3}) reduce to 
\begin{align}
 & \big[ z  (1- z  ) \partial ^2_{z } + (C + \frac{1}{2} - z ) \partial _{z }
 - m^2a^2 \big] h(z )= 0~.
 \label{eq:Gauss_eq}
\end{align}
Here we have dropped suffices $\pm$. The general solution to Eq.(\ref{eq:Gauss_eq}) is given by Gauss's hypergeometric function $F(\alpha , \beta ,\gamma ;z )$ with arbitrary constants $A_{h}$ and $B_{h}$ 
\begin{align}
 & h(\zeta ) = A_{h} F(\alpha , \beta , \gamma ;z) 
  + B_{h} (-z )^{1- \gamma } F(\beta - \gamma +1, \alpha -\gamma +1 ,2-\gamma  ;z) ~,
 \label{eq:solution_u}
\end{align}
where
\begin{align}
 & \alpha = - \beta \equiv  ima~, \quad \gamma \equiv  C + \frac{1}{2}~.
 \label{eq:def_alphe_beta} 
\end{align}
\indent We then consider the conserved current of fermion,
\begin{align}
 & J^{A} = \bar {\psi }\Gamma ^{A} \psi ~,
 \label{eq:def_current}
\end{align}
which satisfies the continuity equation
\begin{align}
 & D_{A}J^{A} = 0~.
 \label{eq:law_current}
\end{align}
In order that the total charge is time-independent, we should require the boundary conditions, which are derived from the integral of Eq. (\ref{eq:law_current}) over some region $V$
\begin{align}
 & \int _{V} d^{6}x\big(\sqrt{-g}D_{A}J^{A}\big) = \int _{V} d^{6}x \partial _{A}\big(\sqrt{-g}J^{A}\big) =0~.
  \label{eq:int_current}
\end{align}
This equation reduces to
\begin{align}
 & \int _{V} d^{6}x \big(\partial _{\theta } \tilde {J}^{\bar {\theta }} +\sin^{-1}{\theta }\partial _{\varphi } \tilde {J}^{\bar {\varphi }} \big)
 =\int _{V} d^{4}x \Big\{\int_{\varphi _1}^{\varphi _2}d\varphi \tilde {J}^{\bar {\theta }}\big|_{\theta _1}^{\theta _2} +\int_{\theta _1}^{\theta _2} d\theta \sin^{-1}{\theta }\tilde {J}^{\bar {\varphi }}\big|_{\varphi _1}^{\varphi _2}\Big\} 
  = 0~.
 \label{eq:int_current2}
\end{align}
where
\begin{align}
 & \tilde {J}^{\bar {A}}=\bar {\tilde {\psi }}\Gamma ^{\bar{A}} \tilde {\psi }~,
\end{align}
$\varphi _1,\varphi _2$ and $\theta _1, \theta_2$ are boundary values. The first and second parts should vanish separately, i.e.,
\begin{align}
 & \tilde {J}^{\bar {\theta }}\big|_{\theta _1}^{\theta _2} =0~, 
 \label{eq:boundary_int1} \\
 & \tilde {J}^{\bar {\varphi }}\big|_{\varphi _1}^{\varphi _2} = 0, 
 \label{eq:boundary_int2}
\end{align}
Explicitly each current is given by
\begin{align}
 & \tilde{J}^{\bar{\theta }} = \bar {\tilde {\psi }}(x^{\mu }, \theta , \varphi )\Gamma ^{\bar{\theta }} \tilde {\psi } (x^{\mu }, \theta , \varphi )
  = i\big[- \bar {\psi}_{R} \psi _{L}\cdot f_{+}^{\dagger}\tau _{1}f_{-} + \bar {\psi}_{L} \psi _{R}\cdot f_{-}^{\dagger}\tau _{1}f_{+}\big]~.
 \label{eq:current_theta_a} 
\end{align}
\begin{align}
 & \tilde{J}^{\bar{\varphi }} = \bar {\tilde {\psi }}(x^{\mu }, \theta , \varphi )\Gamma ^{\bar{\varphi }} \tilde {\psi } (x^{\mu }, \theta , \varphi )
 = i\big[- \bar {\psi}_{R} \psi _{L}\cdot f_{+}^{\dagger}\tau _{2}f_{-}  + \bar {\psi}_{L} \psi _{R}\cdot f_{-}^{\dagger}\tau _{2}f_{+}\big]~.
  \label{eq:current_phi_a}
\end{align}
Since 
\begin{align}
 & f_{+} = g_{+} + g_{-} =  
 \left(
   \begin{array}{c}
       \alpha _{+}(\theta ) u_{+}(\varphi ) + \alpha _{-}(\theta )u_{-}(\varphi )  \\
       \alpha _{+}(\theta )v_{+}(\varphi ) + \alpha _{-}(\theta )v_{-}(\varphi ) \\
   \end{array}
  \right)~,
  \label{eq:cal_f_+0} \\
 & f_{-} = i(g_{+} - g_{-}) = i 
 \left(
   \begin{array}{c}
       \alpha _{+}(\theta )u_{+}(\varphi ) - \alpha _{-}u_{-}(\varphi )  \\
       \alpha _{+}(\theta )v_{+}(\varphi ) - \alpha _{-}v_{-}(\varphi ) \\
   \end{array}
  \right)~,
  \label{eq:cal_f_-0} 
\end{align}
with $\alpha _{\pm}=\tan^{C_{\pm}}\theta /2$, we get $C_{\pm}=0$ from the boundary conditions (\ref{eq:boundary_int1}), that is, $\alpha _{\pm}(\theta )=1$. Hence $f_{\pm}$ become independent of $\theta $, so that we have $\partial _{\theta }\tilde{J}^{\theta }=0$. Since the 4d current conserves, i.e., $\partial _{\mu }\tilde{J}^{\mu }=0$, from the 4d Dirac equation, the $\varphi $ current turns out to be conserved by itself, i.e.,  $\partial _{\varphi }\tilde{J}^{\varphi }=0$. This will be discussed later.\\
\indent The equation (\ref{eq:u_eq2}) with $C_{\pm}=0$ then becomes
\begin{align}
 & \Big[\partial^2 _{\varphi } - \tan{\varphi } \partial _{\varphi }  + \frac{m^2a^2}{\cos^2{\varphi }}\Big] u_{\pm}(\varphi ) 
 = 0~. 
 \label{eq:u_eq2a}
\end{align}
Introducing a variable
\begin{align}
 & w(\varphi ) \equiv  \ln{\frac{1+\sin{\varphi }}{\cos{\varphi }}}=-w(-\varphi )~, \quad w(\frac{\pi}{2}-\varepsilon )\simeq -\ln{\varepsilon }~, \quad w(-\frac{\pi}{2}+\varepsilon ) \simeq \ln{\varepsilon }~,
 \label{eq:def_w} 
\end{align}
where $\varepsilon $ is infinitesimally small positive quantity, Eq.(\ref{eq:u_eq2a}) reduces to
\begin{align}
 & \Big[\partial^2 _{w}  + k^2\Big] u_{\pm}(\varphi ) 
 = 0~, \quad k\equiv ma \neq 0~, 
 \label{eq:u_eq2w}
\end{align}
with a general solution
\begin{align}
 & u_{\pm} =  A_{\pm} e^{ikw(\varphi )} + B_{\pm} e^{-ikw(\varphi )}~,
 \label{eq:solution_u_w}
\end{align}
From Eq. (\ref{eq:u_Eq}) with $C_{\pm}=0$, we have a useful formula between $u_{\pm}$ and $v_{\pm}$ as
\begin{align}
 & v_{\pm} = \pm \frac{1}{k} \partial _{w} u_{\pm}~, \quad k \neq 0~.~,
 \label{eq:solution_v_u}
\end{align}
The same equations (\ref{eq:u_eq2w}) and (\ref{eq:solution_u_w}) are also obtained from Eqs.(\ref{eq:Gauss_eq}) and (\ref{eq:solution_u}) when  $C_{\pm} = 0$. \\
\indent We are now considering a region for $\theta -\varphi $ space by introducing positive small parameters $\varepsilon $ and $\eta $
\begin{align}
 & \varphi _1 \equiv  -\frac{\pi}{2} + \varepsilon \leq  \varphi  \leq \frac{\pi}{2} - \varepsilon  \equiv  \varphi _{2}~,
 \label{eq:region_fai_epsion} \\
 & \theta _{1} \equiv  \eta  \leq  \theta \leq \pi - \eta \equiv  \theta _2~.
 \nonumber
\end{align}
The parameter $\varepsilon $ corresponds with the fact that the fermion wave function cannot be well defined at  $\varphi =\pm \pi /2$. So we consider nearby points $\varphi =\pm(\pi /2-\varepsilon )$. As will be shown later, the parameters $\varepsilon $ and $\eta $ receive a quick response from the fermion back reaction. They will be fixed to be extremely small non-zero quantities, so that, in the regions above, the back reaction from the fermion can be neglected. We regard the other side of $\varphi $, i.e., $\pi /2 \leq \varphi \leq 3\pi/2$, to be a copy of the region $-\pi /2\leq \varphi \leq \pi /2$.\\
\indent Let us then derive the 4D mass formula of fermion. This is nothing but the bound state problem for wave functions $f_{\pm}(\theta ,\varphi )$, which are now independent of $\theta $ and thereby given by
\begin{align}
 & f_{+} = 
 \left(
   \begin{array}{c}
       u_{+}(\varphi ) + u_{-}(\varphi )  \\
       v_{+}(\varphi ) + v_{-}(\varphi ) \\
   \end{array}
  \right)~,
  \label{eq:cal_f_+} \\
 & f_{-} =  i 
 \left(
   \begin{array}{c}
       u_{+}(\varphi ) - u_{-}(\varphi )  \\
       v_{+}(\varphi ) - v_{-}(\varphi ) \\
   \end{array}
  \right)~,
  \nonumber 
\end{align}
Since $u_{\pm}(\varphi )$ satisfy the 2nd order differential equations (\ref{eq:u_eq2a}) or (\ref{eq:u_eq2w}), we require boundary conditions for $u_{\pm}(\varphi )$ as
\begin{align}
 & u_{\pm}|_{\varphi _1, \varphi _{2}} = 0~.
 \label{eq:bc_u_1}
\end{align}
It is not necessary to impose further boundary conditions upon the other component $v_{\pm}$, because  $v_{\pm}$ can be expressed by $u_{\pm}$ through the formula Eqs.(\ref{eq:solution_v_u}). \\
\indent Substituting $u_{\pm} = A_{\pm} \exp{(ikw)} + B_{\pm} \exp{(-ikw)}$ into Eqs.(\ref{eq:bc_u_1}) and using the odd function property $w(-\varphi )=-w(\varphi )=\ln{\varepsilon }$, we have
\begin{align}
 & A_{\pm} e^{ikw} + B_{\pm} e^{-ikw} = 0~,
 \label{eq:cal_bc_1} \\
 & A_{\pm} e^{-ikw} + B_{\pm} e^{ikw} = 0~, 
 \nonumber
\end{align}
to lead
\begin{align}
 & ( A_{\pm} + B_{\pm} ) \cos{(kw)} = 0~, 
 \label{eq:bc_u_2} \\
 & ( A_{\pm} - B_{\pm} ) \sin{(kw)} = 0~.
 \nonumber 
\end{align}
Hence we get
\begin{align}
 & kw=(n-\frac{1}{2})\pi \quad \mbox{when } A_{\pm} = B_{\pm}~, 
 \label{eq:k_condition} \\
 & kw=n\pi \quad \mbox{when } A_{\pm} = -B_{\pm}~, 
 \nonumber
\end{align}
where $w=-\ln{\varepsilon }$, $k=ma$ and $n=1,2,\cdots$, that is,
\begin{align}
 & m=\frac{\pi }{a|\ln{\varepsilon }|}(n-\frac{1}{2})~, \quad \mbox{when } A_{\pm} = B_{\pm}~,
 \label{eq:mass_formula1} \\
 & m=\frac{\pi }{a|\ln{\varepsilon }|}n~, \quad \mbox{when } A_{\pm} = -B_{\pm}~.
 \nonumber
\end{align}
\indent The same mass formula as above can also be obtained by another type of boundary conditions
\begin{align}
 & v_{\pm}|_{\varphi _1, \varphi _{2}} = \pm \frac{1}{k} \partial _{w} u_{\pm}|_{\varphi _1, \varphi _{2}}= 0~.
 \label{eq:bc_u_3}
\end{align}
more generally, by any linear combination between them, i. e.,
\begin{align}
 & \alpha u_{\pm} + \beta \partial _{w} u_{\pm}|_{\varphi _1, \varphi _{2}} =  0~.
 \label{eq:bc_u_4}
\end{align}
However, we should discard a type of $f_{+}|_{\varphi _1, \varphi _{2}} =0$ (or $f_{-}|_{\varphi _1, \varphi _{2}} =0$). This gives the same mass formula as above, but we are led to a meaningless result $A_{\pm}=B_{\pm}=0$  from the constraints  $H=0$ (\ref{eq:condition_H}) below. \\
\indent The parameters $\varepsilon $ and $\eta $ are given by
\begin{align}
 & |\varphi | \leq \frac{\pi}{2} - \varepsilon ~, \quad \eta \leq \theta \leq \pi - \eta ~,
 \label{eq:region_epsilon_theta_a} \\
 & \varepsilon = \big(\frac{a}{a_0}\big)^{1/3}~, \quad \eta = \big(\frac{a}{a_0}\big)^{1/4}~, 
 \nonumber
\end{align}
where $a$ is the radius of internal 2D sphere with a unit length $a_{0}$  made $\varepsilon , \eta $  dimensionless. Here we have assumed to be $a<<a_0$. These inequalities come from the inequality
\begin{align}
 & \big|T_{AB}^{(b)}\big| >> \big| T_{AB}^{(f)}\big|~,
 \label{eq:ineualities_T} 
\end{align}
where $T_{AB}^{(b)}$ is the bulk energy-momentum tensor given by Eq.(\ref{eq:T_AB2}), while $T_{AB}^{(f)}$ is the fermion energy-momentum tensor. If the above inequality holds, the back reaction from the fermion may be neglected. The detail will be discussed in the Appendix. \\ 
\indent Substituting  $\varepsilon =(a/a_{0})^{1/3}$ into Eq.(\ref{eq:mass_formula1}) we get the 4D fermion mass formula for non zero-mode 
\begin{align}
 & m=\frac{3\pi }{a|\ln{(a/a_{0})}|}l~,
 \label{eq:mass_formula3}
\end{align}
where $l$ takes positive integral or half-odd integral values. This formula is nothing but for the KK modes. However, as seen in the Appendix the fermion mass formula will be invalid for so large values $l$, because it is approaching to the 6D Planck mass.\\
\indent Finally let us discuss the $\varphi $ current conservation $\partial _{\varphi } \tilde {J}^{\varphi }=0$ by itself. Not that
\begin{align}
 & \tilde{J}^{\bar{\varphi }}= \bar {\tilde {\psi }}(x^{\mu },\theta , \varphi )\Gamma ^{\bar{\varphi }} \tilde {\psi }(x^{\mu },\theta , \varphi )
 = i\big[-\bar{\psi }_{R}\psi _{L}f_{+}^{\dagger}\tau _{2}f_{-} + \bar{\psi }_{L}\psi _{R}f_{-}^{\dagger}\tau _{2}f_{+} \big]~, 
 \label{eq:cal_current_epsilon}
\end{align}
where
\begin{align}
 & f_{+}^{\dagger}\tau _{2}f_{-} = (u_{+}^{*}v_{+} - v_{+}^{*}u_{+}) - (u_{-}^{*}v_{-} - v_{-}^{*}u_{-})
- (u_{+}^{*}v_{-} - v_{+}^{*}u_{-}) + (u_{-}^{*}v_{+} - v_{-}^{*}u_{+})~.
 \label{eq:f_f2a}
\end{align}
Here we have dropped the $\theta $ factor because of $|\alpha (\theta )|^2=1$. We can use the formulas (\ref{eq:solution_v_u}) and $u_{\pm}=A_{\pm}\exp{(ikw)}+B_{\pm}\exp{(-ikw)}$ to get
\begin{align}
 & f_{+}^{\dagger}\tau _{2}f_{-} = 2i\big[|A_{+}|^2 - |B_{+}|^2 + |A_{-}|^2 - |B_{-}|^2
 \label{eq:f_f2b} \\
 &-A_{+}^{*}B_{-} e^{-2ikw} + B_{+}^{*} A_{-} e^{2ikw} + A_{+}B_{-}^{*} e^{2ikw} - B_{+} A_{-}^{*} e^{-2ikw}\big]~.
 \nonumber
\end{align}
The current $J^{\varphi }$, therefore, reduces to
\begin{align}
 & \tilde{J}^{\bar{\varphi }}= 2\big(\bar{\psi }_{R}\psi _{L}-\bar{\psi }_{L}\psi _{R}\big)\big(He^{2ikw}  - H^{*}e^{-2ikw} \big) + 2 \big(\bar{\psi }_{R}\psi _{L}+\bar{\psi }_{L}\psi _{R}\big) I~,
 \label{eq:cal_current_epsilon2}
\end{align}
where
\begin{align}
 & H = B_{+}^{*}A_{-} + B_{-}^{*}A_{+}~. \quad I = |A_{+}|^2 - |B_{+}|^2 + |A_{-}|^2 - |B_{-}|^2~.
 \label{eq:def_H_I}
\end{align}
From the continuity equation $\partial _{\varphi }\tilde{J}^{\varphi }=0$, we have $\exp{(2ikw)}H + \exp{(-2ikw)}H^{*}=0$. Since this equation holds for any $w$, we should have a constraint
\begin{align}
 & H = B_{+}^{*}A_{-} + B_{-}^{*}A_{+}=0~.
 \label{eq:condition_H}
\end{align} 
%
\subsection{The zero mode solution in 4-dimensions}
%
We can put $m=0$ in Eqs.(\ref{eq:f_pm_Eq}) to lead equations
\begin{align}
 & \hat{\nabla } f_{\pm}(\theta , \varphi ) = \Big(\tau _{1} \partial _{\theta } + \tau _{2} \frac{1}{\sin{\theta }}\partial _{\varphi }\Big) f_{\pm}(\theta , \varphi )= 0~.
 \label{eq:f_pm_Eq0} 
\end{align}
Separating variables by
\begin{align}
 &  f_{\pm} (\theta , \varphi ) 
 =  \alpha_{\pm} (\theta ) \left(
   \begin{array}{c}
      a_{\pm} \\
      b_{\pm} \\
   \end{array}
  \right)~,
 \label{eq:solution_f_pm_0} 
\end{align}
we get
\begin{align}
 & \partial _{\theta } \alpha (\theta ) =  \frac{C_{\pm}}{\sin{\theta } } \alpha (\theta ) ~,
 \label{eq:alpha_Eq0} \\
 & \partial _{\varphi } a_{\pm}(\varphi ) = iC a_{\pm}(\varphi ) ~,
 \label{eq:a_Eq} \\
 & \partial _{\varphi } b_{\pm}(\varphi ) = -iC b_{\pm}(\varphi ) ~,
 \label{eq:b_Eq}
\end{align}
General solutions to these equations are
\begin{align}
 & \alpha (\theta ) =  \tan^{C_{\pm}}{\frac{\theta }{2} }~,
 \label{eq:alpha_Eq0a} \\
 & a_{\pm}(\varphi ) = P_{\pm} e^{iC_{\pm}\varphi } ~,
 \label{eq:a_Eq1} \\
 & b_{\pm}(\varphi ) = Q_{\pm} e^{-iC_{\pm}\varphi } ~,
 \label{eq:b_Eq1}
\end{align}
where $P_{\pm}$ and $Q_{\pm}$ are some constants. Substituting these solutions into the current boundary conditions (\ref{eq:boundary_int1}) we get again $C_{\pm}=0$. Then the wave functions $f_{\pm}(\theta , \varphi )$ become constants, satisfying the boundary conditions (\ref{eq:boundary_int2}) automatically. \\
\indent To sum up, we have the zero mode constant solution ( $m=0$)
\begin{align}
 &  f_{\pm} (\theta , \varphi ) 
 =  \alpha_{\pm} (\theta ) \left(
   \begin{array}{c}
      a_{\pm} \\
      b_{\pm} \\
   \end{array}
  \right)
  =  \left(
   \begin{array}{c}
      P_{\pm} \\
      Q_{\pm} \\
   \end{array}
  \right)~,
 \label{eq:solution_f_pm_1} 
\end{align}
where $P_{\pm}$ and $Q_{\pm}$ are some constants, subjecting to normalizations (regions of variables are given by Eqs. (\ref{eq:region_fai_epsion})).
%
%
\section{Concluding remarks} 
%
In the 6D warped world with the extra 2D surface of a sphere, we have proposed an exactly solvable model for fermion masses. We have derived the finite mass formula for 4D fermion (\ref{eq:mass_formula3}), with the zero mode ($m=0$)  solution (\ref{eq:solution_f_pm_1}).\\
\indent The warp factor is given $\phi (\theta ,\varphi )=\sin{\theta }\cos{\varphi }$, which is a solution to Einstein's equation with the bulk cosmological constant $\Lambda $ and the energy-momentum tensor (\ref{eq:ansatz}) of the bulk matter fields. Hence our metric has zeros of the warp factor at $\varphi =\pm\pi/2$  and $\theta =0, \pi$. The internal wave functions then cannot be well defined at these zero points $\varphi =\pm\pi/2$. Therefore, we have put boundary conditions (4.19), (4.20) and also (4.32) not at $\varphi =\pm\pi/2$, but at nearby points $\varphi =\pm(\pi/2-\varepsilon )$.\\
\indent The mass formulas have been obtained from such modified boundary conditions with parameters $\varepsilon $ and $\eta $, which are fixed from the requirement that the back reaction of the fermion should be neglected. Namely, in the regions
\begin{align}
 & -\frac{\pi }{2}+\varepsilon \leq  \varphi  \leq \frac{\pi }{2} -\varepsilon ~,\quad \eta \leq \theta \leq \pi -\eta~, 
 \label{eq:regions}
\end{align}
the back reaction of the fermion should be neglected. These parameters are given by $\varepsilon = (a/a_{0})^{1/3}$ and $\eta = (a/a_{0})^{1/4}$, where $a$ is the radius of the extra 2D sphere with a unit length $a_0$. \\
\indent The internal function has a form of $u=e^{iC\varphi }h$. When $C$ is real and takes half-integer values,  $u$ is  $4\pi$-periodic, regarded as a spinor on the sphere. However, this possibility fails,   $C$ eventually becomes zero from the condition (4.19). The internal function $u$ cannot be the $4\pi$-periodic spinor on the sphere, but the function $\psi (x)$ is the spinor in the four-dimensional space \cite{ref:Gogberashvili1}. \\
\indent The normalization integral for non-zero mode wave functions is given by
\begin{align}
 & I = \int d^3x d\theta d\varphi \sqrt{-g}\psi ^{\dagger}\psi = \int d^5x \big[ \psi _{R}^{\dagger}\psi_{R}f_{+}^{\dagger}f_{+} + \psi _{L}^{\dagger}\psi_{L}f_{-}^{\dagger}f_{-} \big]~,
 \label{eq:normalization} \\
 & f_{\pm}^{\dagger}f_{\pm}= 2\big[|A_{+}|^2 + |A_{-}|^2 + |B_{+}|^2 + |B_{-}|^2 \pm \big( He^{2ikw} + H^{*}e^{-2ikw} \big)\big]~.
  \nonumber
\end{align}
Since $H=0$ according to the constraint (\ref{eq:condition_H}), then the integral is convergent.\\
\indent Our model provides another possibility of obtaining fermion zero mode, rather than traditional model based on Dirac's monopole.

\section*{Acknowledgments}

We would like to express our deep gratitude to T. Okamura for many valuable discussions.

\appendix
\section{Constraints from the back reaction of fermion}\label{sec:appendA}
%
The fermion energy-momentum tensor is given by
\begin{align}
 & T_{AB}^{\ (f)} \equiv  \frac{i}{2} \bar {\psi} \big(\Gamma _{A}\overrightarrow {D}_{B} - \overleftarrow {D}_{B}\Gamma _{A}\big)\psi  - g_{AB}  \frac{i}{2} \bar {\psi} \big(\Gamma ^{M}\overrightarrow {D}_{M}-\overleftarrow {D}_{M}\Gamma ^{M}\big)\psi~,
  \label{eq:T_MN} 
\end{align}
where the second term becomes zero for the 6D massless fermion. Substituting $\psi =(\phi ^{2}\sin^{1/2}{\theta })^{-1}\tilde {\psi }$ into Eq.(A.1) we have for the $\mu \nu $ component
\begin{align}
 & T_{\mu \nu }^{(f)} = \frac{i}{2\phi^3 \sin{\theta }} \bar {\tilde {\psi}} \big(\Gamma _{\bar{\mu }}\overrightarrow {D}_{\bar{\nu }} - \overleftarrow {D}_{\bar{\nu }}\Gamma _{\bar{\mu }}\big)\tilde {\psi}
 \label{eq:T_mu_nu} \\
 & = \frac{i}{2\phi^3 \sin{\theta }}\big\{\bar {\tilde {\psi}} \big(\Gamma _{\bar{\mu }}\overrightarrow {\partial }_{\bar{\nu }}-\overleftarrow {\partial }_{\bar{\mu }}\Gamma _{\bar{\nu }} - \frac{\cos{\theta }\cos{\varphi }}{2a}\big[\Gamma _{\bar{\mu }},~\Gamma _{\bar{\nu }}\big]\Gamma _{\bar{\theta }} 
 + \frac{\sin{\varphi }}{2a}\big[\Gamma _{\bar{\mu }},~\Gamma _{\bar{\nu }}\big]\Gamma _{\bar{\varphi }} \big)\tilde {\psi}  \big\}
 \nonumber \\
 & = \frac{1}{2\phi^3 \sin{\theta }}\big\{\tilde{t}_{\bar{\mu} \bar{\nu }} - \frac{\cos{\theta }\cos{\varphi }}{2a}\tilde{s}_{\bar{\mu }\bar{\nu }}^{\theta } + \frac{\sin{\varphi }}{2a}\tilde{s}_{\bar{\mu }\bar{\nu }}^{\varphi } \big\}~,
 \nonumber  
\end{align}
where
\begin{align}
 & \tilde{t}_{\bar{\mu }\bar{\nu }}\equiv i\bar {\psi_R} (\gamma _{\mu }\overrightarrow {\partial _{\nu }} - \overleftarrow {\partial }_{\nu }\gamma _{\mu })\psi_R\cdot f_{+}^{\dagger}f_{+}
 + i\bar {\psi_L} (\gamma _{\mu }\overrightarrow {\partial _{\nu }} - \overleftarrow {\partial }_{\nu }\gamma _{\mu })\psi_L\cdot f_{-}^{\dagger}f_{-}~.
 \label{eq:C_mu_nu} \\
& \tilde{s}_{\bar{\mu }\bar{\nu }}^{\theta } \equiv \bar {\psi_R} \big[\gamma _{\bar{\mu }},~\gamma _{\bar{\nu }}\big]\psi_L\cdot f_{+}^{\dagger}\tau _{1}f_{-}
 - \bar {\psi_L} \big[\gamma _{\bar{\mu }},~\gamma _{\bar{\nu }}\big]\psi_R\cdot f_{-}^{\dagger}\tau _{1}f_{+}~. 
 \nonumber \\
& \tilde{s}_{\bar{\mu }\bar{\nu }}^{\varphi } \equiv \bar {\psi_R} \big[\gamma _{\bar{\mu }},~\gamma _{\bar{\nu }}\big]\psi_L\cdot f_{+}^{\dagger}\tau _{2}f_{-}
 - \bar {\psi_L} \big[\gamma _{\bar{\mu }},~\gamma _{\bar{\nu }}\big]\psi_R\cdot f_{-}^{\dagger}\tau _{2}f_{+}~. 
 \nonumber 
\end{align}
Then from the inequality (\ref{eq:ineualities_T}) we are enough to check only for diagonal parts
\begin{align}
 & \frac{3\Lambda }{10} >> \frac{1}{|2\phi^3 \sin{\theta }|}|\tilde{t}_{\bar{\mu }\bar{\mu }}|~, 
 \label{eq:condition1}
\end{align}
which reduces to
\begin{align}
 & |\sin^4{\theta } \cos^3{\varphi }| >> \frac{5}{3\Lambda }|\tilde{t}_{\bar{\mu }\bar{\mu }}|~.
 \label{eq:condition2}
\end{align}
Since the 4D energy-momentum tensor element $\tilde{t}_{\mu \mu }$ may be extremely smaller than the 6D Planck mass, i.e.
\begin{align}
 & |\tilde{t}_{\bar{\mu }\bar{\mu }}| << \frac{1}{\kappa_{6} ^2 a_{0}^2}~,
 \label{eq:Planckmass}
\end{align}
where $a_{0}$  is a unit length with the inequality $a/a_0 << 1$. This reduces to
\begin{align}
 & \frac{5|\tilde{t}_{\bar{\mu }\bar{\mu }}|}{3\Lambda } <<  \frac{5}{3\Lambda \kappa_{6}^2 a_{0}^2}=\frac{1}{6}\big(\frac{a}{a_{0}}\big)^2~,
 \label{eq:Planclmass2}
\end{align}
by using Eq.(\ref{eq:solution_Fai}). Hence we get inequalities
\begin{align}
 & |\sin^4{\theta }\cos^3{\varphi }| \geq \frac{1}{6}\big(\frac{a}{a_0}\big)^2>> \frac{5}{3\Lambda } |\tilde{t}_{\bar{\mu }\bar{\mu }}| ~.
 \label{eq:condition2a}
\end{align}
\indent For another component we have
\begin{align}
 & T_{55}^{(f)} = \frac{i}{2} \bar {\psi}  \big(\Gamma _{\theta }\overrightarrow {D}_{\theta } - \overleftarrow {D}_{\theta }\Gamma _{\theta }\big)\psi
 = \frac{i}{2\phi^4 \sin{\theta }} \bar {\tilde {\psi}}  \big(\Gamma _{\bar{\theta }}\overrightarrow {\partial }_{\bar{\theta }} - \overleftarrow {\partial }_{\bar{\theta }}\Gamma _{\bar{\theta }}\big)\tilde {\psi}
 ~,
 \label{eq:T_55}
\end{align}
From Eq.(\ref{eq:alpha_Eq}) with $C_{\pm}=0$, we have
\begin{align}
 & T_{55}^{(f)} = 0~,
 \label{eq:T_55_A}
\end{align}
Hence Eq.(\ref{eq:T_55_A}) is satisfied with the inequality (\ref{eq:ineualities_T})\\
\indent For the 66 component we get
\begin{align}
 & T_{66}^{(f)} =\frac{i}{2} \bar {\psi} \big(\Gamma _{\varphi }\overrightarrow {D}_{\varphi } - \overleftarrow {D}_{\varphi }\Gamma _{\varphi }\big)\psi 
  = i\frac{a}{2\phi^4 }\bar {\tilde {\psi}} \big(\Gamma _{\bar{\varphi }}\overrightarrow {\partial }_{\bar{\varphi }} - \overleftarrow {\partial }_{\bar{\varphi }}\Gamma _{\bar{\varphi }}\big)\tilde {\psi}~,
\label{eq:T_66_2}
\end{align} 
Using Eq.(\ref{eq:f_pm_Eq}) we get
\begin{align}
 & T_{66}^{(f)}   =\frac{ma^2}{\phi^4 \cos{\varphi }} N ~,
\label{eq:T_66_2_A}
\end{align} 
where $N=\frac{1}{2}(\bar {\psi_R} \psi_L+\bar {\psi_L} \psi_R)(f_{+}^{\dagger}f_{+} + f_{-}^{\dagger}f_{-})\sim O(1)$. \\
Near $\varphi =\pm \pi /2$, the inequality becomes
\begin{align}
 & \frac{6a^2\Lambda \sin^2{\theta }}{10\phi^2} >> \frac{ma^2|N|}{\phi^4 |\cos{\varphi }|}~. 
 \label{eq:condition5a}
\end{align}
This reduces to
\begin{align}
 & |\sin^4{\theta } \cos^3{\varphi }| >> \frac{10m|N|}{6\Lambda }=\frac{m|N|a^2\kappa_{6} ^2}{6}~.
 \label{eq:condition6}   
\end{align}
Since the energy term $m|N|$ is considered to be extreamly smaller than the 6D Planck mass ($1/\kappa_{6} ^2 a_{0}^2$), we have
\begin{align}
 & \frac{1}{\kappa_{6} ^2 a_{0}^2} >> m|N|~,
 \nonumber
\end{align}
that is, 
\begin{align}
 & 1>> \kappa_{6} ^2 a_{0}^2 m|N|\equiv \lambda ~,
 \label{eq:condition_lambda}
\end{align}
Substituting this into (\ref{eq:condition6}) we get
\begin{align}
 & |\sin^4{\theta } \cos^3{\varphi }| \geq \frac{1}{6}\big(\frac{a}{a_{0}}\big)^2 >> \frac{\lambda }{6}\big(\frac{a}{a_{0}}\big)^2 ~.
 \label{eq:condition7}   
\end{align}
This is the same form as Eq.(\ref{eq:condition2a}). Henceforth we neglect the facter (1/6) in both equations.\\
\indent Since $a/a_{0}<<1$, the region of $|\sin^4{\theta }|\times |\cos^3{\varphi }|$ may be approximated by the square region of $|\sin^4{\theta }|\geq (a/a_{0})$ and $|\cos^3{\varphi }|\geq (a/a_{0})$. Therefore we reach a conclusion that 
\begin{align}
 & |\varphi | \leq \frac{\pi }{2} -\varepsilon ~, \quad \varepsilon = (a/a_{0})^{\frac{1}{3}}.
 \label{eq:boundary_phi} \\
 & \eta \leq \theta \leq \pi -\eta ~, \quad \eta = (a/a_{0})^{\frac{1}{4}}~.
 \label{eq:regions_theta}
\end{align}
From Eq.(\ref{eq:condition_lambda}) we see that the fermion mass formula will be invalid for so large values $l$, because it is approaching to the 6D Planck mass.


\end{document}